# Dynamic coupling between the LID and NMP domain motions in the catalytic conversion of ATP and AMP to ADP by adenylate kinase

Biman Jana, Bharat V. Adkar, Rajib Biswas and Biman Bagchi*

Solid State and Structural Chemistry Unit

Indian Institute of Science, Bangalore 560012, India

*E-mail: bbagchi@sscu.iisc.ernet.in

## ABSTRACT

The catalytic conversion of ATP and AMP to ADP by adenylate kinase (ADK) involves large amplitude, ligand induced domain motions, involving the opening and the closing of LID and NMP domains, during the repeated catalytic cycle. We discover and analyze an interesting dynamical coupling between the motions of the two domains during the opening, using large scale atomistic molecular dynamics trajectory analysis, covariance analysis and multi-dimensional free energy calculations *with explicit water*. Initially, the LID domain must open by a certain amount before the NMP domain can begin to open. Dynamical correlation map shows interesting cross-peak between LID and NMP domain which suggests the presence of correlated motion between them. This is also reflected in our calculated two dimensional free energy surface contour diagram which has an interesting *elliptic shape*, revealing a strong correlation between the opening of the LID domain and that of the NMP domain. Our free energy surface of the LID domain motion is rugged due to interaction with water and the signature of ruggedness is evident in the observed RMSD variation and its fluctuation time correlation functions. We develop a correlated dynamical disorder type theoretical model to explain the observed dynamic **coupling between the motions of the two domains in ADK. Our model correctly reproduces several features of the cross-correlation observed in simulations.**

## I. INTRODUCTION

Understanding a protein's function at an atomic level has been revolutionized by high-resolution X-ray crystallography, resulting in a surge in the studies of structure–function relationships. However, too much reliance on these structures can be deceptive, suggesting that one unique structure, the 'folded state', contains all the final answers. The dynamic nature of biology might play a role at the microscopic level. Recent experiments show that protein/ enzyme is a dynamic entity which samples a large ensemble of conformations around the average structure as a result of fluctuations driven by thermal energy[1-3]. Thus, a complete description of a given protein's function may require understanding of a multidimensional free energy landscape that defines the relative probabilities of the conformational states and the energy barriers between them. In biology, this concept has recently gained importance, leading to an extension of the structure–function paradigm to include dynamics. To understand proteins in action, the fourth dimension, the time, must be added to the snapshots of proteins frozen in X-ray crystal structures[1-3]. A major obstacle is that it is not possible to watch experimentally individual atoms moving within a protein. The role of these fluctuations in enzyme activity is still unclear.

Some recent studies on the role of conformational fluctuation in the biological activity of a few enzymes revealed that the attainment of certain critical conformation is essential for the chemical step/ reaction to occur[4-7]. These studies also clearly suggest that this conformational cycling step need not always be the rate determining step. A non-equilibrium theory of the enzymatic reaction has been formulated where it has been discussed how an enzyme can operate in a non-equilibrium steady state to enhance the rate of catalysis[6-7]. This study emphasized the point that by staying in a non-equilibrium steady state, the enzyme can minimize the free energy barrier required for conformational displacement[6-7]. Paradoxically, operating from this non-equilibrium steady state may actually lead to a *reduction of the role of conformational fluctuations in catalysis*. However, this theory assumes the existence of an intermediate state in the relaxation of the enzyme conformation *after the product release and prior to another substrate capture*[6-7].

Experiments and several simulation studies have indeed revealed existence of large scale conformational fluctuations (domain motion) in the catalytic conversion of $Mg^{2+}$-ATP + AMP → $Mg^{2+}$-ADP + ADP by the enzyme adenylate kinase (ADK)[1-3, 8-11]. Our recent simulation study of the enzyme with explicit water indicated the existence of a half-open-half-closed (HOHC) intermediate state of the enzyme which accelerates the rate of catalysis with the reduction of the barrier of conformational cycling[12].

Despite considerable studies[1-3, 8-15], the precise sequence of events in the catalysis and the role of ADK conformational fluctuations in the reaction process are not well understood. The following questions need to be answered.

(1) Is there any correlation between the motion of LID and NMP domains during opening?

(2) Is one of the domains open first and induce the movement of the other? Which one opens first and why?

(3) What is the reason for such a correlation, if it exists?

(4) How does this correlation enhance the rate/efficiency of the catalysis?

In this study, we use large scale atomistic molecular dynamics trajectory analysis and multi-dimensional free energy calculations *with explicit water* to investigate the dynamical coupling between the motions of the two domains during the opening. The main results of our present study are as follows.

(1) The LID domain must open first to certain amount before the NMP domain begins to open.

(2) Our computed two dimensional free energy surface contour diagram has an interesting *elliptic shape*, revealing a strong correlation between the conformational motions of the LID domain and that of the NMP domain.

(3) The free energy surface of the LID domain motion is found to be rugged due to interaction with water and the signature of ruggedness is evident in the observed RMSD variation and its time correlation functions.

(4) Our free energy surface of the NMP domain is smooth with a larger harmonic confining force constant (near the minimum). This is the reason for smaller fluctuations of the NMP domain.

(5) We also find a clear signature of the allosteric correlation between the motions of the two domains in the free energy surface of the NMP domain motion.

(6)  Finally, we develop a dynamical disorder based theoretical model, but with correlations, to describe the observed dynamic coupling between the motions of the two domains in ADK. Our model correctly reproduces the features of cross-correlation observed in simulations.

Onuchic *et. al.*[9] have earlier investigated the conformational transition in adenylate kinase using nonlinear normal mode analysis and calculated the strain energy associated with different possible paths of the catalytic cycle. Many of the results of their analysis are in agreement with our present analysis, although emphasis of the present study is entirely different.

The organization of the rest of the paper is as follows. In the next section, simulation details are discussed. The conformational fluctuations in the two domains of ADK and the dynamical correlation between their fluctuations are discussed in section III and section IV, respectively. We discuss the dynamical correlation between the LID and NMP domain motion in terms of covariance analysis in section V. Signatures of dynamical coupling in the free energy surface for the domain motions are discussed in section VI. We discuss the role of dynamical coupling in the catalysis in section VII. A theoretical model to explain the observed dynamical coupling is presented in section VIII before concluding the paper in section IX.

## II. SIMULATION DETAILS

**Fig. 1.** shows the open and closed forms of the adenylate kinase superimposed on each other. The enzyme can be viewed as consisting of following three domains. (a) The LID domain that closes on the ATP binding to the ADK. (b) The NMP domain that closes on the AMP binding to the ADK, and (c) the CORE domain which shows no significant conformational changes upon ligand binding. The enzyme changes its conformation from open state to closed

state during the ligand binding event. The LID and NMP domains again open up to release the product and get ready to capture a new set of substrates for next round of catalysis.

We have characterized the two conformational motions by measuring two distances. (1) The distance between the centers of mass of the LID and the CORE domains, denoted by $R^{CM}_{LID-CORE}$ and (2) the distance between the centers of mass of the NMP and the CORE domains, denoted by $R^{CM}_{NMP-CORE}$. The values of the $R^{CM}_{LID-CORE}$ for open and closed forms are 29.5 Å and 20.5 Å, respectively, and the same for the $R^{CM}_{NMP-CORE}$ for open and closed forms are 21.0 Å and 18.3 Å, respectively[16,17]. The LID domain is defined as residues 118-160, the NMP domain as residues 30-67, and the CORE domain as residues 1-29, 68-117, and 161-214.

As mentioned earlier, several studies have been carried out to understand the role of conformational fluctuations on catalysis, both experimentally and computationally. Here we report for the first time a long time scale (individual runs of maximum of 500 ns and total run time 2.5 μs), a fully atomistic simulation of ADK in explicit solvent environment. The long time scale simulation provides us a unique opportunity to understand the possible role of conformational fluctuations and sequence of binding events that might be happening in ADK.

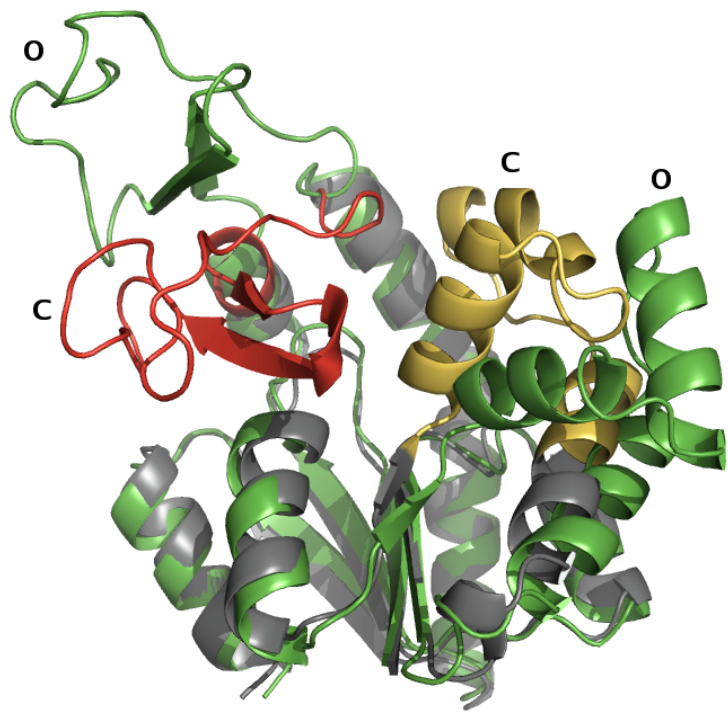

**FIG. 1. Ribbon diagram showing the open and the closed forms of the adenylate kinase.** Superposition of closed (1AKE[17]) and open (4AKE[16]) forms of adenylate kinase. The open form is colored green. The LID domain in closed form is colored in red, the NMP domain is in yellow, and the CORE domain is in gray.

The x-ray structure of the open form (4AKE[16]) was used as the starting structure for open state simulations. To generate the starting structure for the closed state simulation, ligand (inhibitor AP5) coordinates were removed from the x-ray structure of the closed form (1AKE[17]). All crystallographic water molecules were stripped off from the pdb files. GROMACS v3.3.1 suite of programs was used for molecular dynamics and other structural analysis[18]. Proteins were centered in a cubic box of 77.3 Å. The box size was so chosen that no atom in the protein, neither in the open nor in the closed conformation, will be less than 10 Å from any of the box-boundaries. All atom topologies for proteins were generated with the help of pdb2gmx and AMBER94 set of parameters (available through AMBER port for GROMACS). The proteins were solvated with pre-equilibrated SPC/E water model[19] using genbox. Total of 14511 and 14496 water molecules were added to closed and open state boxes, respectively. Four sodium atoms were added to achieve electrically neutrality. The systems were put through the following energy minimization and equilibration steps: 1) Steepest descent energy minimization, 2) a 100 ps of position-restrained NPT simulation with restraining force constant of 1000 kJ.mol$^{-1}$nm$^{-2}$, 3) a 100 ps of NVT equilibration without restraints, 4) a 100 ps of NPT equilibration, 5) the final NPT simulation of 350 ns. During simulations, temperature was maintained at 300 K using Nose-Hoover thermostat with $\tau_T$ = 0.1 ps and pressure was maintained isotropically at 1 bar using Berendsen barostat with $\tau_P$ = 0.5 ps. Isothermal compressibility of water was set to $4.5\times10^{-5}$ bar$^{-1}$. The trajectories are built with structures at 2 ps interval.

**RMSD:** Root mean square deviation for backbone was calculated with the help of g_rms program when the backbone of CORE domain was used for superimposition. The structures from both the trajectories were superimposed onto the closed conformation of ADK. A common reference for both the trajectories was chosen so that the distribution of RMSD can be easily studied. LID domain is defined as residues 118-160, NMP as 30-67, and CORE domain is defined as residues 1-29, 68-117, and 161-214.

**Angle and distance:** The Cα atoms of following residues were used to calculate distance between LID and CORE domains with g_dist: Val148 (from LID) and Glu22 (from CORE). The angle calculation was done by considering Cα atom of Pro9 (from hinge) in addition to Cα from Val148 and Glu22 with g_angle.

**Umbrella sampling:** While $R^{CM}_{LID-CORE}$ varies from 20.5 Å to 29.5 Å while going from closed conformation to open conformation, $R^{CM}_{NMP-CORE}$ varies from 18.3 Å to 21.5 Å. A two-dimensional free-energy surface was constructed by taking $R^{CM}$ as reaction coordinates. $R^{CM}_{LID-CORE}$ varied from 19.0 Å to 32 Å, and $R^{CM}_{NMP-CORE}$ from 17 Å to 23 Å, both at an interval of 0.5 Å. Force constant of 3000 kJ.mol$^{-1}$.nm$^{-2}$ was used to restrain the domains at respective $R^{CM}$. Umbrella sampling simulations were performed using pull code from GROMACS v4.0.5. Free energy surfaces are extracted using wham program. The simulations involved more than 80,000 water molecules for the free energy calculations. We have performed 2 ns long simulation for each umbrella window. We have used a convergence criterion of 0.0001 in successive $F_j$ values.

**Correlation analysis:** Cross- and auto-correlation of RMSD evolution was calculated for LID and NMP domains by following definition: $C_x(\Delta t) = \langle C_p(0) \bullet C_q(t) \rangle$. For auto-correlation ($x$ = auto), $p$ and $q$ were either LID or NMP. For cross-correlation ($x$ = cross), $p$ and $q$ were set to LID and NMP, respectively. The maximum correlation time considered was half of the simulation length.

**Covariance analysis:** We analyzed the presence of dynamic correlation between LID and NMP openings. In covariance analysis, a correlation map is derived from $C_\alpha$ fluctuations during the trajectory started from closed state. Correlations in atomic fluctuations were computed using the generalized linear correlation coefficient which is based on the linear mutual information parameter $I_{lin}$[20].

$$I_{lin}(x_i, x_j) = \frac{1}{2}\left[\ln \det C_{(i)} + \ln \det C_{(j)} - \ln \det C_{(ij)}\right]$$

where,

$$C_i = \langle x_i^T x_i \rangle,$$

and

$$C_{(ij)} = \langle (x_i, x_j)^T (x_i, x_j) \rangle$$

are the marginal covariance of atom *i*, and the pair-covariance matrix of atom *i* and *j*, respectively. The covariance parameters were calculated for $C_\alpha$ of all residues after aligning the backbone of CORE domain and taking the closed state as reference state. Thus, this analysis can provide clear information about the correlation between the motion of different regions of a protein. We have used this method instead of Pearson coefficient analysis because the estimate of correlations from the Pearson coefficient are only strictly valid if $x_i$ and $x_j$ are colinear vectors, as already pointed out by Ichiye and Karplus[21]. However, the present method based on linear mutual information parameter does not impose such a restriction on the atom motion.[20]

## III. INTRINSIC CONFORMATIONAL FLUCTUATION AND OCCASIONAL JUMPS IN THE LID MOTION

ADK needs to undergo intrinsic large amplitude motions during the conformational cycle. These fluctuations may be affected by the presence of the ligand, but the enzyme, even in the free state should be capable of executing conformational fluctuations. Both LID and the NMP domains need to move by a considerable amount to allow the ATP and AMP to get in and then again open to let the products (the two ADP molecules) go out. A key step in the catalysis (in addition, of course, to the actual chemical reaction step of the phosphate transfer) can be another substrate capture immediately after the product release so that the catalytic cycle can continue. We have examined the intrinsic conformational fluctuation of ADK by following the variation of root mean square deviation (RMSD), inter-domain distances and hinge angles. We have plotted all these three quantities for the LID domain in **Fig. 2**. LID domain motion has an interesting character of caging and followed by hopping motion similar to glassy dynamics[22]. It fluctuates around an average LID-CORE distance for quite some time (30 -50 ns) (caging dynamics) and then jumps to another LID-CORE distance within 5-10 ns (jump dynamics) and fluctuates around the new average LID-CORE distance again for quite a some time. In analogy with the glassy dynamics, it can be implied that the free energy surface of the LID domain motion is quite rugged with several metastable minima and barriers[23]. This implied ruggedness of the free energy surface of the LID domain motion and its effect on the nature of self and cross-correlation function of the RMSD fluctuation involving LID domain are discussed later. However, we find a very smooth but slow progress (completes approximately by 50 – 60 ns) of the NMP domain towards the open state indicating relatively smooth free energy surface. This analysis also suggests that the conformational fluctuation in ADK is mainly due to the inter-domain displacement and not due to the change in structure of corresponding domains. Therefore, we choose the centre of mass distances between LID and CORE ($R^{CM}_{LID-CORE}$) and NMP and CORE ($R^{CM}_{NMP-CORE}$) as the two order parameters for the free energy surface calculation.

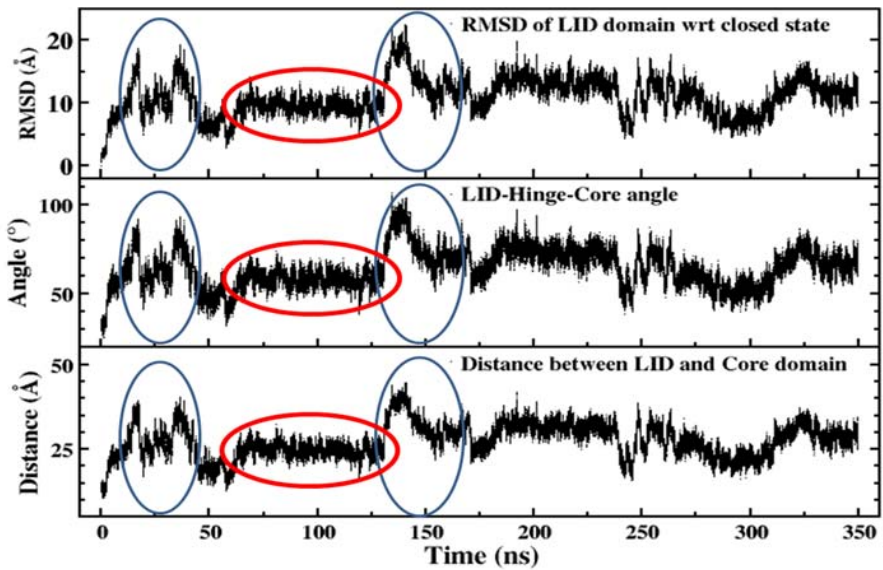

**FIG. 2 Comparison of fluctuations to confirm the inter-domain nature of the movements.** The RMSD of LID domain with respect to the native structure of the closed state, LID-hinge-CORE angle, and LID-CORE distance distribution as a function of simulation time for closed state simulation. The Cαs of following residues are used to calculate distance between LID and CORE domains: Val148 (from LID) and Glu22 (from CORE). The angle calculation is done by considering Cα atom of Pro9 (from hinge) in addition to Cα from Val148 and Glu22. Note the occasional jumps and practically invariant value of the quantities in the highlighted regions. Note also the similarity in the fluctuations of the different quantities as highlighted by the selected region of ellipses. The region highlighted by the red ellipse shows the fluctuation in the HOHC state.

If the domains themselves undergo change in structures during the simulation, it will contribute to alterations in RMSD. To check this possibility, we calculate the radius of gyration ($R_G$) of all the three domains separately and also LID-CORE and NMP-CORE inter-domain combinations. It is well-known that $R_G$ is a good measure of the compactness of the structure. If any of the domains undergoes change in its structure due to the unfolding/ cracking, then the $R_G$ of that domain will increase. **Figure 3** shows the $R_G$ of intra- and inter-domain motion calculated from the closed state simulation. We find no significant change of $R_G$ of the individual domains over time, indicating that the individual domains do not melt or unfold significantly over time. However, the variation of inter-domain $R_G$ is found to be similar to the RMSD fluctuation of the corresponding domain. This again indicates the dominant role of inter-domain motion of the conformational fluctuation in ADK.

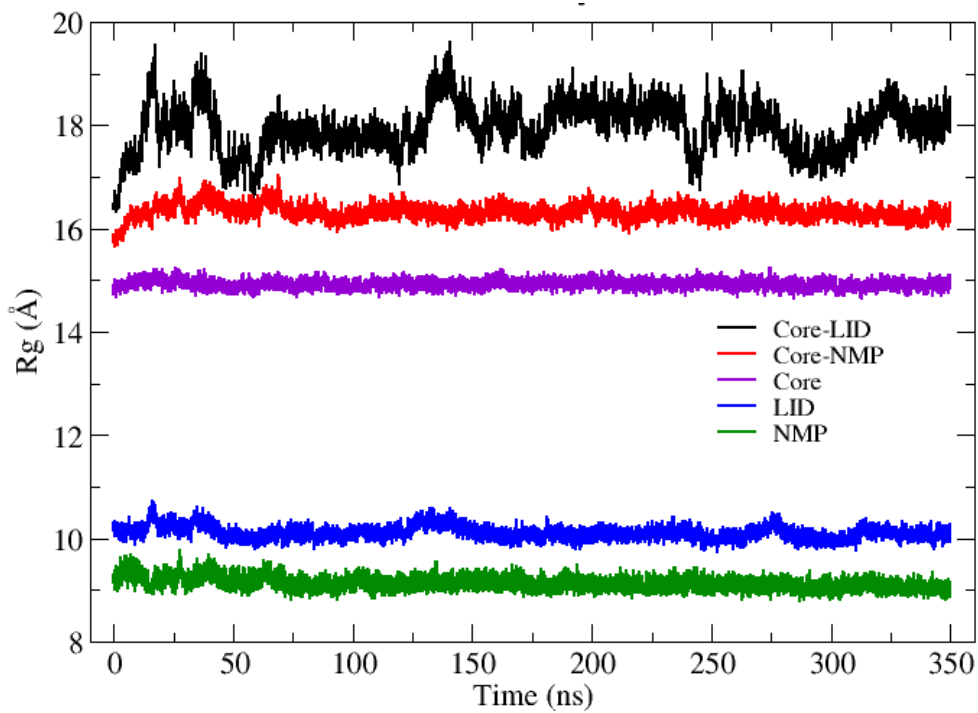

**Figure 3:** **Radius of gyration ($R_G$) as a function of simulation time is calculated for various domains.** $R_G^{CORE}$ is shown in magenta color, $R_G^{LID}$ in blue, and $R_G^{NMP}$ in green. Additionally, $R_G$ for LID and NMP with CORE domain is also calculated and shown in black ($R_G^{CORE-LID}$) and blue ($R_G^{CORE-NMP}$) colors respectively. Note the invariant nature of the $R_G$ throughout the whole simulation period for the individual domains indicating that the individual domains do not melt or unfold significantly.

# IV. DYNAMICAL CORRELATION BETWEEN MOTIONS OF LID AND NMP DOMAINS

The correlation in the motion of the two domains during the opening from the closed state is an important issue that remains to be understood. To this end, we have analyzed carefully the RMSD fluctuation of the LID and NMP domains at the start of the closed state simulation. We performed this simulation by removing the inhibitor from the closed state (**1AKE**) and then allowed the system to evolve to its open state. RMSD fluctuations of both the domains are presented in **Fig. 4.** We found that initially LID domain started opening faster than the NMP domain and reached at an intermediate state by ~12 ns, with pronounced signatures of rugged landscape, discussed later. The NMP domain started to open slowly after the initial opening of the LID domain and went on to the open state by 50 ns, smoothly (signature of smooth landscape). Such correlation between the fluctuations in RMSD of the two domains motivated us to calculate self and cross-correlation function of the RMSD fluctuations of the domains. Self and cross-correlation (definitions are provided in the **Method** section) functions are presented in **Fig. 5**. We found a fast decay of the self-correlation in the RMSD fluctuation of the LID domain within 20-25 ns and a relatively slow decay of the self-correlation function for the NMP domain, complete by 75 - 80 ns. There are several other interesting features in the decay pattern of the cross-correlation function of the RMSD fluctuations of the two domains. Initially the correlation decreased up to 20 ns and then it increased between 20-30 ns followed by a final decay to zero by ~ 50 ns. Thus, at the start of the domain opening, the information of the LID domain opening gets transmitted to the NMP domain approximately within 20 ns and then NMP domain also starts opening with the LID domain leading to an increase in the cross-correlation between 20-30 ns. The cross-correlation decayed to zero by 50 ns which was the time needed to NMP domain to reach the open state. These results suggest that the opening of LID domain induces the opening of the NMP domain and there exists no correlation once the domains reach a sufficiently open conformation. Such correlations are important to understand the allostery between the domain motions which is an important aspect of the catalysis[24-25]. Earlier studies have completely missed such a correlation in the domain motion in ADK as total simulation time was not enough for those studies[1, 25].

We found an interesting pattern in the decay of the RMSD fluctuation time correlation functions. The time correlation functions involving the RMSD fluctuation of the LID domain (LID self and cross) showed occasional increase (jumps) in the correlation. However, the self time correlation function of the RMSD fluctuation of NMP domain showed no such jumps in the decay pattern. Such jumps in the correlation function bear the signature of the presence of multiple maxima and minima along the opening motion of the LID domain.

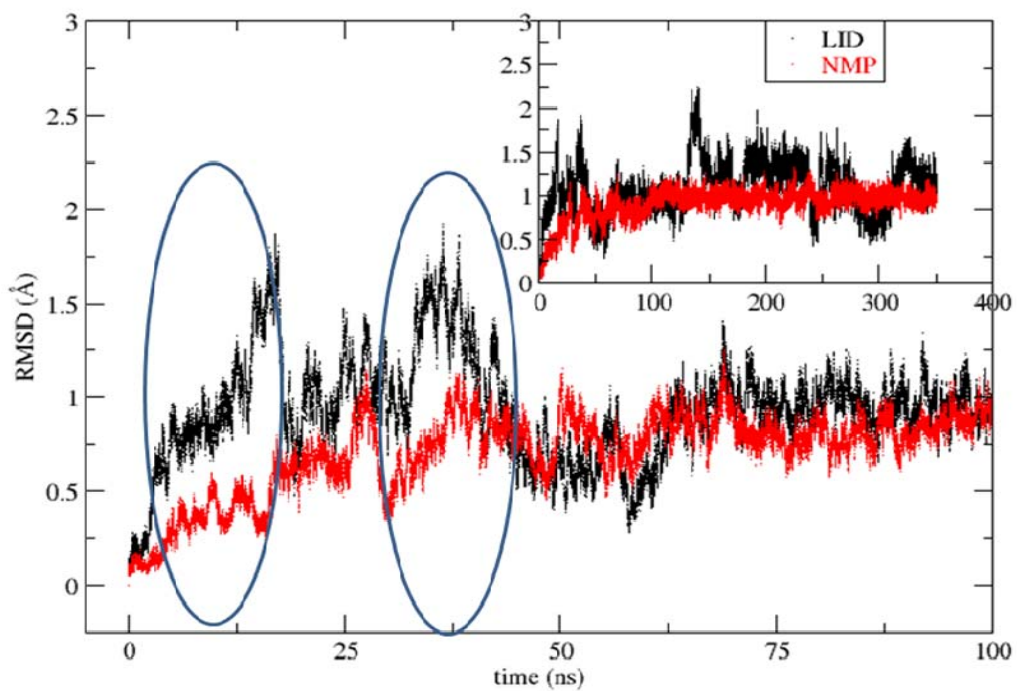

**FIG. 4. Correlation between the movement of the two domains of ADK during opening.** Time evolution of the RMSD of the LID and NMP domains during the closed state simulation. The selected regions show the correlations in their fluctuations while opening.

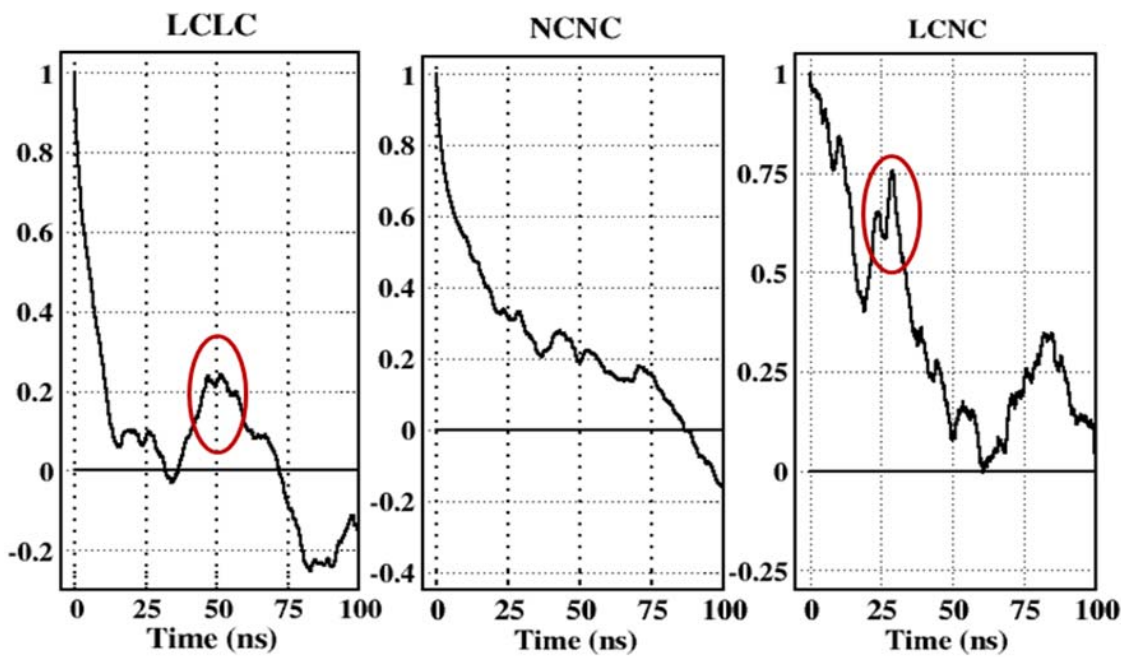

**FIG. 5. Self- and cross correlation functions of the RMSD fluctuation between the two domains.** Note the faster and slower initial decay of the self-correlation functions of the fluctuation of LID and NMP domains, respectively. Note also the occassional increase (highlighed with red ellipses) of the correlation function involving LID domain motion indicating a rugged free energy landscape.

## V. COVARIANCE ANALYSIS

In order to confirm the correlation between the motion of the two domains, we have performed covariance analysis[20]. This method is used regularly to explore the correlated motion between the different distant regions in the protein/enzyme[20]. We analyzed the dynamic correlation map derived from the fluctuation of $C_\alpha$ atoms of the enzyme (details of the calculation is described in the method section). The correlation map is shown in **Fig. 6**.

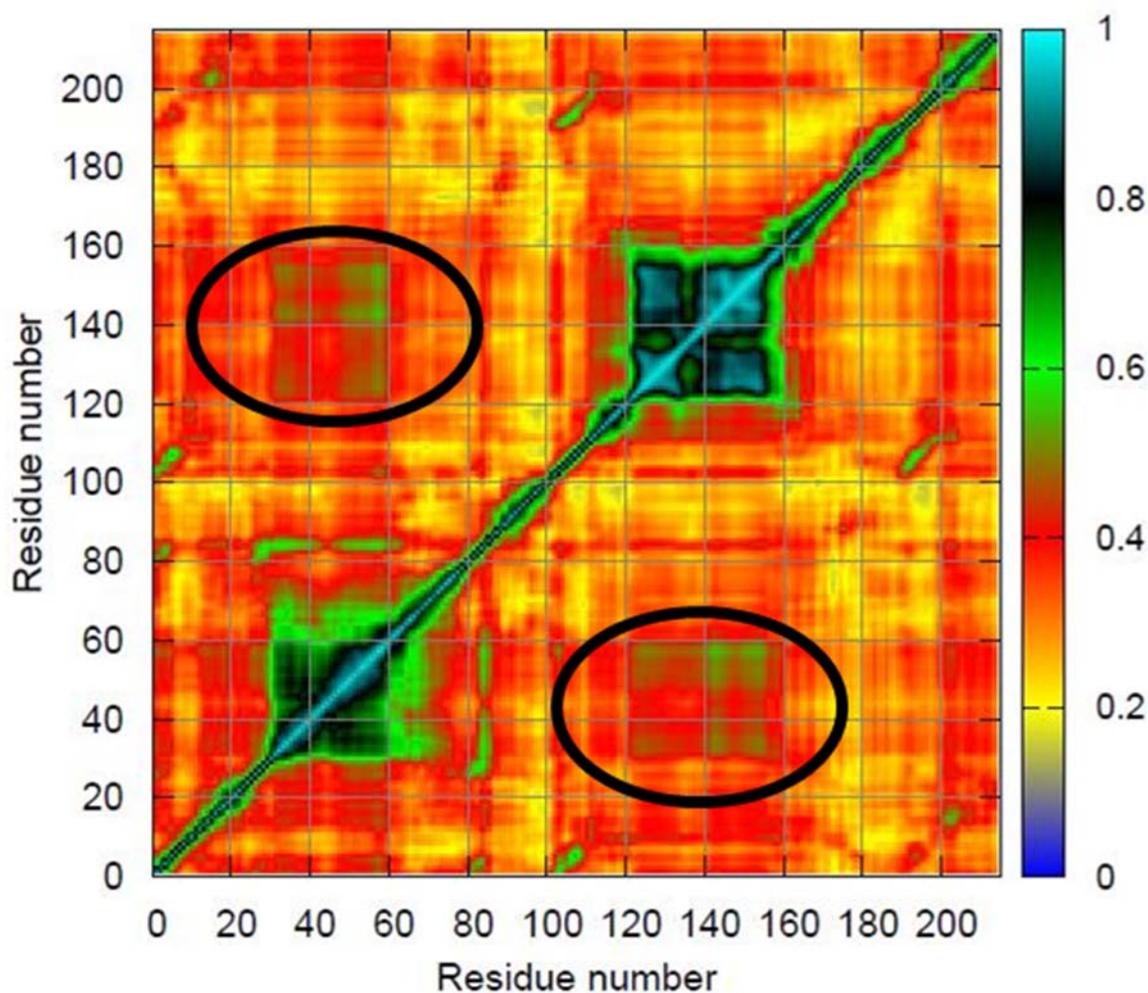

**FIG. 6.** Dynamic correlation map derived from $C_\alpha$ fluctuation in adenylate kinase. The cross peak highlighted by the ellipse is due to the correlated motion between LID and NMP domains (two peaks highlighted here are identical). Two regions along the diagonal line indicate the highly correlated motion inside the domains.

We find an interesting cross peak corresponding to LID and NMP domains (two peaks highlighted by ellipses are identical as the correlation matrix is symmetric) which has reasonably higher intensity. This indicates a strong correlation between the motion of the two domains as discussed in the earlier section.

Therefore, the correlation observed in the earlier section has now further been confirmed by this analysis. Additionally, we observed two regions along the diagonal line corresponding to LID and NMP domains separately. These two peak regions indicate a coherently correlated motion of all the atoms in a domain. This observation also supports inter domain nature of the fluctuation in adenylate kinase with out any significant unfurling in respective domains.

# VI. THE TWO-DIMENSIONAL ELLIPTIC FREE ENERGY SURFACE AND SIGNATURE OF DYNAMIC COUPLING

The free energy surface of the corresponding domain motion of the ADK was calculated by using umbrella sampling (values of the parameter used are provided in the **Method** section). The free energy surface was obtained for the LID domain motion at *13 distances* of NMP-CORE and for the NMP domain motion at *27 distances* of LID-CORE.

The two-dimensional free energy surface of the free enzyme is presented in **Fig. 7a**. The conformations of ADK with $R^{CM}_{LID-CORE} \approx 26$ Å and $R^{CM}_{NMP-CORE} \approx 19$ Å form a stable minimum for the free enzyme. This conformation belongs neither to the fully open state nor to the fully closed state ensemble. This ensemble of stable intermediate conformation is termed as *half-open-half-closed (HOHC)* state hereafter. The presence of such a stable intermediate state is further confirmed by the trajectory analysis. We have also found that the existance of this HOHC state modifies the catalytic cycle of ADK which runs between HOHC and closed state rather than between open and closed state. This results are discussed discussed elsewhere[12].

An important aspect of this two dimensional free energy surface is its *elliptic nature, with the LID-CORE distance as the major axis*. This indicates that the LID domain motion is less constrained than that of the NMP domain. Thus, both the opening and the closing of the enzyme proceed via the opening and the closing LID domain as it has to climb relatively less barrier. This also suggests that the motion of the LID domain can facilitate the similar motion in NMP domain if the LID-NMP interface interaction plays crucial role in stabilizing the closed state. Indeed the importance of this interface interaction in the closed state has already been discussed in the literature[10,11]. This elliptic nature of the free energy surface implies a relatively more labile LID domain motion and this can play crucial role in the prediction of the sequence of events in the catalytic cycle and also in the prevention of the misligation[10]. This issue will be discussed later.

**(a)**

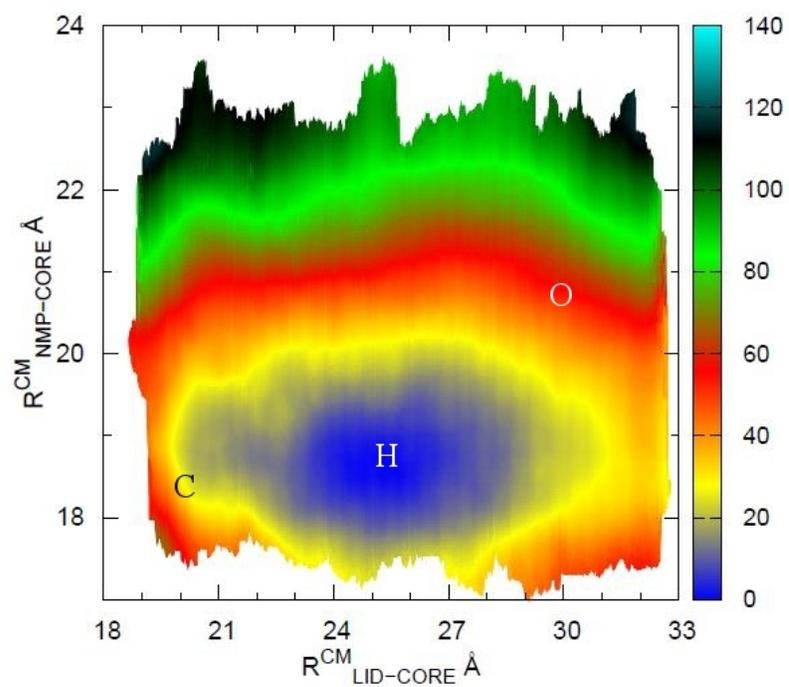

C = Closed state, H = Half-open-half-closed state and O = Open state

**(b)**

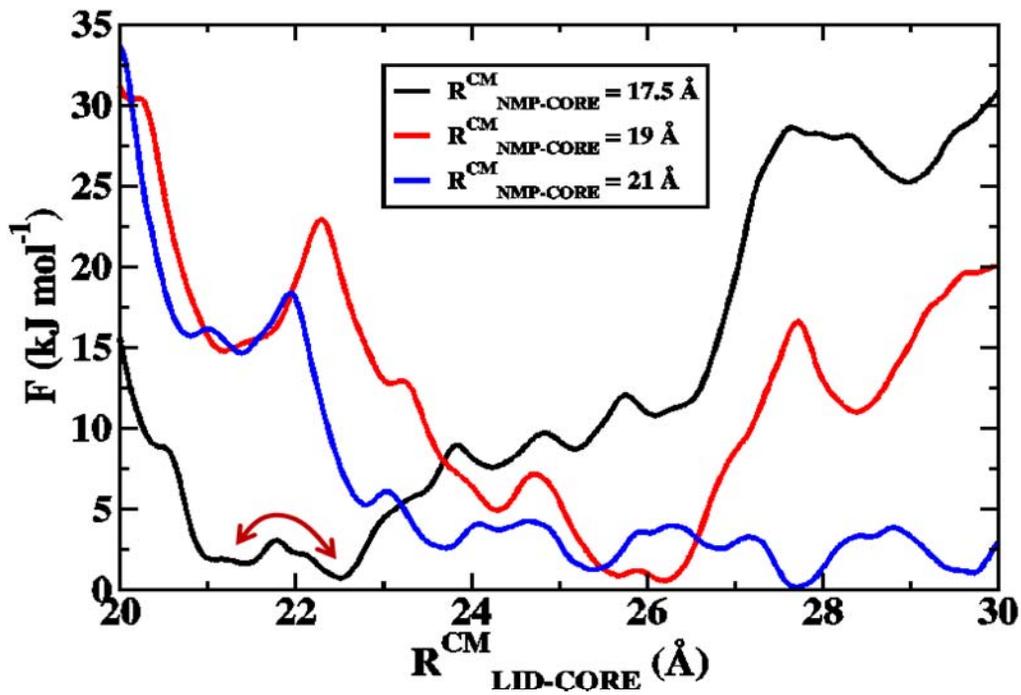

**(c)**

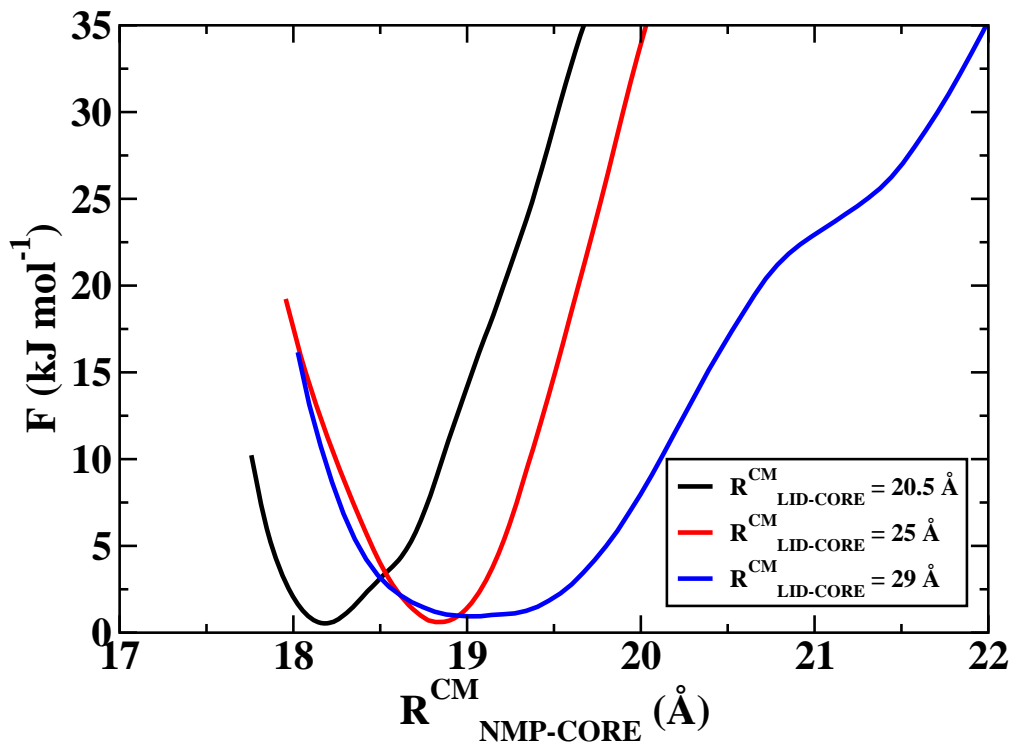

**FIG. 6. (a) Two-dimensional free energy surface of the free ADK enzyme.** Two dimensions are the center of mass distances of the LID and NMP domains from the CORE. The color code has been so chosen that the closely spaced regions can be distinguished clearly. Note the elliptic nature of the free energy surface. **(b) Free energy landscape of the LID domain motion.** Free energy surface calculated from umbrella sampling for the LID domain motion at three different $R^{CM}_{NMP-CORE}$ distances, 17.5 Å (black), 19 Å (red) and 21 Å (blue). Note the flat and wide minima when the NMP domain in closed and open. Note also the rugged landscape of all the free energy surfaces. **(c) Free energy landscape of the NMP domain motion.** Free energy surface calculated from umbrella sampling for the NMP domain motion at there different $R^{CM}_{LID-CORE}$ distances, 20.5 Å (black), 25 Å (red) and 29 Å (blue). Note that with increasing opening of the LID domain the free energy minimum is shifting towards the open state and the surface is becoming increasingly less steep.

## VI.1 Rugged free energy surface of the LID domain motion

In **Fig. 7b,** we show the free energy surfaces of the LID motion for three NMP-CORE separations ($R^{CM}_{NMP-CORE}$). When the NMP domain was kept near its closed conformation ($R^{CM}_{NMP-CORE}$ = 17.5 Å), a rugged but flat minima in the wide range of LID-CORE distance (starting from closed to HOHC state for LID) was obtained. This is consistent with the observed elliptic nature of the two dimensional free energy surface discussed earlier. LID can fluctuate in the wide range of LID-CORE distances while NMP is in the closed state. Opening of the domains can start with the LID motion easily. While opening its LID domain, ADK first gets trapped in the minima around 21 Å. It can then move to the minima around 22.5 Å because these two states are separated by very low free energy barrier. However, it has to climb a sufficiently large barrier (~12 kJ/mol) to get into the next flat minima at larger opening of the LID domain (24 to 26 Å). Thus it stays in that conformation (21 to 22.5 Å) for quite some time. This kind of feature for the LID domain motion is observed in the trajectory, as discussed earlier. For an intermediate $R^{CM}_{NMP-CORE}$ (~ 19 Å), we find a stable minimum for an intermediate state of LID domain which has a $R^{CM}_{LID-CORE}$ around 26 Å. Interestingly, when NMP domain is near the fully open state ($R^{CM}_{NMP-CORE}$ = 21 Å), we again find rugged but flat free energy minima in the wide range of LID-CORE distances (starting from HOHC state to the fully open state of the LID domain). This implies that the closing of the domains can start with the LID motion easily. These results of the relatively easy opening and closing of the LID domain compared to NMP is in agreement with earlier studies[10,11]. Note that all the free energy y surfaces of LID domain motion are rugged. Due to the easy movement of LID domain, it can sample a wide range of conformational space.

In this process the domain will make transitions between several stable intermediates which are separated by little barrier. The occurrence of several intermediate states is primarily due the increase in water mediated interaction upon opening of the domain[12]. This introduces the ruggedness in the surface. This ruggedness introduces the occasional jumps and the signature of caging dynamics in the trajectory of the LID domain motion.

## VI.2 Relatively steeper free energy surface of the NMP domain motion

**Fig. 7c.** displays the free energy surfaces of the NMP domain motion for three $R^{CM}_{LID-CORE}$ separations. For a fully closed LID domain ($R^{CM}_{LID-CORE}$ = 20.5 Å), the free energy surface of the NMP domain motion exhibited a stable minimum in the closed state ($R^{CM}_{NMP-CORE}$ = 18.1 Å) and increase from this minimum was found to be relatively steep. This implied that the NMP domain opening needs the opening of the LID domain. Such a scenario indicated an allosteric correlation between the domain motions[24,25]. When the LID domain was in the intermediate state ($R^{CM}_{LID-CORE}$ = 25 Å), we find that the minimum is shifted to $R^{CM}_{NMP-CORE}$ = 18.8 Å indicating the opening of the NMP domain with partial opening of the LID domain. For a fully open LID domain ($R^{CM}_{LID-CORE}$ = 29.5 Å), the free energy surface exhibited a nearly flat minimum starting from 18.5 to 19.8 Å. Thus, NMP free energy surfaces showed that it opens gradually (because of the larger free energy cost for conformational motion of the NMP domain). However, the surface becomes increasingly flat with increasing opening of the LID domain. This relatively steeper free energy surface of the NMP motion when the LID domain in closed state and the continuous sojourn of the NMP domain towards the open state with increasingly flat surface as the LID domain opens are in agreement with the earlier studies[10,11]. This can also play a crucial role in the catalytic cycle of ADK as will be discussed later.

# VII. POSSIBLE ROLE OF DYNAMICAL COUPLING ON THE CATALYSIS

There are two types of coupling possible for the domain motions in the ADK enzyme. (1) In one of the scenarios, the conformational change of the enzyme occurs between the static structures due to the ligand binding. The conformational change in one domain due to ligand binding signals the initiation of the same in another domain and eventually transition between two enzyme structures occurs. This is called the ligand induced allosteric coupling. (2) In the other scenario, the enzyme can fluctuates between different intermediate structures in the absence of ligand as enzyme is a dynamical entity. The dynamical coupling associated with the conformational changes in ligand free condition is called intrinsic allosteric coupling. In the present paper, we have explored the intrinsic dynamical coupling in the conformational fluctuation of ADK.

Results obtained from the above analysis suggested that the LID domain fluctuates easily compared to the NMP domain during both opening and closing. The initial movement of the LID domain also triggers the similar movement in the NMP domain. Thus, it is convenient to start the process of domain opening (closing) with the LID domain followed by the NMP domain as this path require little barrier to climb to start with. In a catalytic cycle, this intrinsic fluctuation and coupling determine the path of the process initially and then ligand induced coupling takes over and completes the process. Thus in the present case, we suggest the possible sequence of events for a complete cycle based on our results as follows.

(1) In the unligated conformation of ADK, LID domain starts closing process initially and subsequently ATP binds to it.

(2) The LID domain closing induces the closing process in the NMP domain with concomitant binding of the AMP.

(3) Phosphoryl transfer occurs resulting in two bound ADP's.

(4) The LID domain starts opening fist and subsequently ADP gets released from it.

(5) The opening of the LID domain induces the opening of the NMP domain with concomitant release of the ADP from it.

(6) Both LID and NMP domains start their respective sojourn towards the open states, but this is slow in the case of LID domain because of the ruggedness of the free energy surface. It reached only up to the HOHC state where it grabs a new substrate and starts to close again.

Next, we suggest the possible role of dynamical coupling on the prevention of the misligation which can give rise to non-productive substrate bound complex giving rise to a lower efficiency. The product forming complex of the ADK is $LID_C^{ATP}$-$NMP_C^{AMP}$. Let us now consider the conformation of ADK in which both the domains are open. In this situation, it has been observed experimentally that the binding affinities of ATP favorably discriminate over AMP by 3kcal/mol[27] and the complex $LID_C^{ATP}$-$NMP_O$ can accommodate only AMP (neither ATP nor ADP) in the NMP domain. Dynamical coupling suggests that the initial step is the closing of LID domain over the ATP (due to binding affinity discrimination) bound to that domain and it eventually give rise to the product forming complex. If the alternative closing path is selected, then AMP binds to the NMP domain to form $LID_O$-$NMP_C^{AMP}$. As we have discussed earlier, the closing of NMP domain needs the LID to close, indicating a relatively faster closer of the LID domain to stabilize the closed state of NMP. In this process LID can close without a ligand or may be with an incorrect ligand (AMP) which gives rise to non-product forming complex. The selection of the correct path is necessary for the high efficiency of the ADK catalysis and this selection is governed by the dynamical coupling.

In case of the domain opening after the product formation, the LID domain starts opening first with the concomitant release of the ADP. Opening of the LID domain signals the opening in the NMP domain due to dynamical coupling and the second ADP gets released. However, if the alternate path is chosen, then NMP domain will open first and ADP gets released from it. In such a situation the LID domain can open and release the ADP or the NMP domain can close immediately with accommodating one AMP to form an unwanted $LID_C^{ADP}$-$NMP_C^{AMP}$. Thus, in helping the selection of the correct path for both opening and closing of the domains, the dynamical coupling (discovered here) helps prevention of misligation.

## VIII. CORRELATED DYNAMICAL DISORDER MODEL

An elegant theoretical model of the propagation of correlation between different parts of protein was developed by Thompson and coworkers[28-30]. These authors employed an Ising model to account for the propagation of interaction between two distant domains (for example between the effector site and the active site of the enzyme). This model could describe the emergence of positive and negative cooperatively[28-30]. We discuss here a somewhat modified adaptation of this model to explain the observed correlation between LID and NMP domains.

In our Ising model description, a molecular domain is represented by a spin which can have two states – spin up (+) or spin down (-). The interaction between two nearest neighbor spins is taken as ferromagnetic. Let us consider that in the absence of a ligand, one particular state at the receptor side (say, the open state) is preferred. This is reproduced by the presence of an external magnetic field on the spins initially to keep them up (spin up represents the open state). When the ligand comes, the opposite spin state is preferred and this is described by switching the sign of the field for that particular site. This change now propagates along the ring, and tries to create a different polarization at a distant spin due to ferromagnetic nearest neighbor interaction. Creation of the opposite polarization (spin down) at the distant site is tantamount to making that site active.

Therefore, in our model, we have a competition between the ferromagnetic interaction which tries to make the neighboring spins parallel and this propagates through the chain and the magnetic field at the active site, as explained in more detail below. When at t=0, we turn the spin at the one end (state at the LID or in more general case, the state at the effector site) upside down, the information propagates through chain interaction and ultimately this promotes the spin down state at the distant site (the NMP site here or the active site in more general scenario). Thus, this competes with the magnetic field on that spin which tries to keep the spin in the up state (at the active site). The present model bears the similarity with some of the dynamical disorder models[31-32] well-known in the literature. Here the disorder comes from the freedom of motion of the spins.

In the present case, we start with the condition that both the LID and NMP domains are closed. These are created by external magnetic fields at both the sites. At t = 0, the constrain that LID domain is closed is removed. In the present context, the extra stability of the HOHC state is maintained by placing a magnetic field in the direction opposite to the one used to create the closed state. Thus, this is quite similar to the presence/absence of a ligand at the receptor site in allosteric enzymes.

At the simplest level, we consider a four spin system, to describe the LID, the first connector (hinge 1), the NMP domain and the second connector (hinge 2). The model is described schematically in **Fig. 8**. The one dimensional Ising Hamiltonian with periodic boundary condition is

$$H = -J\sum_i \sigma_i \sigma_{i+1} - \sum_i h_i \sigma_i ,$$

where the first sum is over all the nearest neighbors. In our correlated dynamical disorder model, we have placed *unequal opposite nonzero* external magnetic field for spin-1 and spin-3.

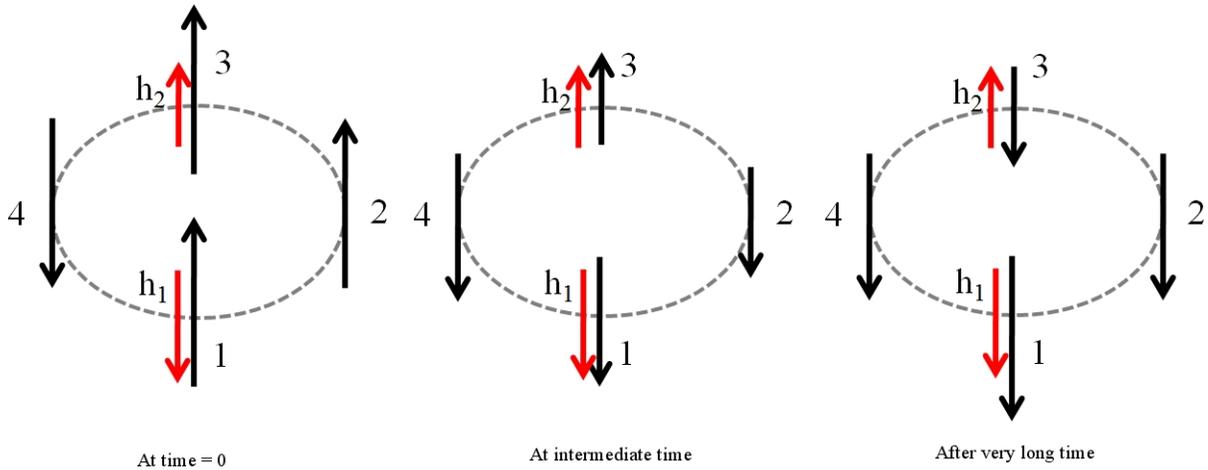

**FIG. 8.** Schematic representation of the model for propagation of dynamical correlation developed in the present study. The black arrows showing the direction of spin polarization at each side and the red arrows are indicating the direction of magnetic field. The time evolution of the spin polarization at each side is shown here.

We have performed Monte Carlo simulation using metropolis algorithm[33], in the presence of external magnetic fields (the direction of the magnetic field at spin 1 is down and at spin 3 is up), with ferromagnetic coupling between the spins. In **Fig. 9**, we plot the cross time correlation function between the spin 1 and 3. We have calculated the cross-correlation function for all the 16 initial configurations possible for the four spin system and the average cross-correlation function calculated using weighted Boltzmann averaging. Here the correlation time is represented by the number of Monte Carlo steps. Initially the correlation decreases to an intermediate time followed by an increase in their correlation. Such a pattern is in agreement with the behavior observed in the cross correlation function between the RMSD fluctuation of LID and NMP domain of ADK plotted in **Fig. 4**. This implies that the information of the conformational change in the LID domain takes certain time to propagate to the NMP domain (intermediate time where cross-correlation function starts to increase). This is a signature of signal transduction between the different domains of the enzyme. Our present model captures this phenomenon correctly. This model presented here can be extended in different directions, for example, to include many spins and also one can allow different degree of correlations between the adjacent spins.

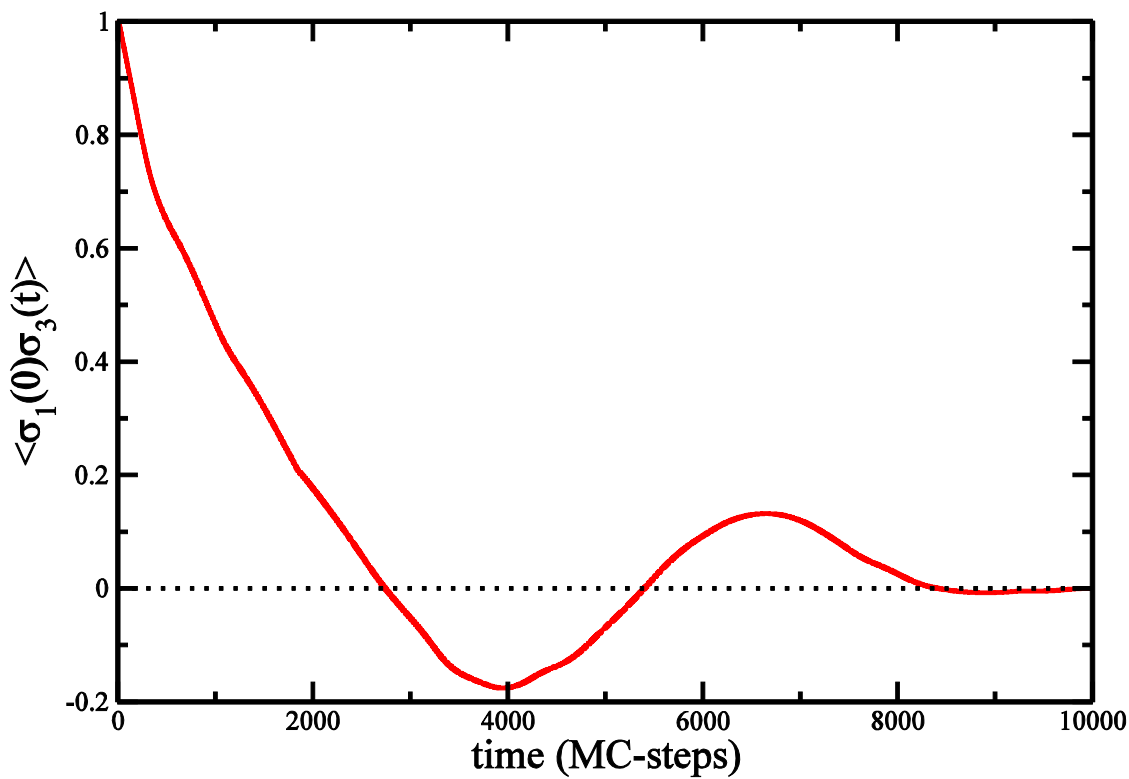

**FIG. 9.** Cross correlation function between spin 1 and 3. Note the initial decrease and the increase of the correlation after an intermediate time. The time here is represented by the number of MC steps.

## IX. CONCLUSION

Let us first summarize the main results of this work. We used large scale atomistic molecular dynamics trajectory analysis and free energy calculations with explicit water to study the catalytic conversion of $Mg^{2+}$-ATP + AMP $\rightarrow$ $Mg^{2+}$-ADP + ADP by adenylate kinase (ADK). This reaction involves large amplitude domain motions, involving the opening and the closing of LID and NMP domains, during the repeated catalytic cycle. We have studied the intrinsic fluctuations, that is, motions in the absence of any ligand. We discovered an interesting dynamical coupling between the motions of the two domains during the opening using covariance analysis of the trajectory. Study of a two dimensional free energy surface and dynamic cross-correlation revealed that initially the LID domain must open to certain degree before the NMP domain can begin to open. The free energy surface of the LID domain motion is rugged due to interaction of polar residues with water and the signature of ruggedness is evident in the observed RMSD variation and its fluctuation time correlation functions. The free energy surface of the NMP domain is steeper and much smoother than that of the LID. We developed an Ising model type dynamic disorder model to explain the observed dynamic coupling between the motions of the two domains in ADK.

Future work will concentrate on the detailed origin of the dynamic coupling and its possible role in catalysis. We also plan to extend the present dynamic disorder model of allosteric regulation.

**ACKNOWLEDGEMENTS**

This work was supported in parts by grants from DST (India) and CSIR (India). BB was supported by a J.C. Bose Fellowship. BJ and RB thank CSIR for a Fellowship.